\documentclass[superscriptaddress,preprintnumbers,amsmath,amssymb,twocolumn,floats,aps,prb]{revtex4-1}
\usepackage{txfonts}
\usepackage{amssymb}
\usepackage{graphicx}
\usepackage{epstopdf}
\usepackage{sidecap}
\usepackage{CJK}
\usepackage{txfonts}
\usepackage{url}

\begin{document}

\bibliographystyle{unsrt}
\title{Photoemission Spectroscopic Evidence of Multiple Dirac Cones in Superconducting BaSn$_3$}

\author{Z. Huang}
\thanks{Equal contributions}
\affiliation{Center for Excellence in Superconducting Electronics, State Key Laboratory of Functional Materials for Informatics, Shanghai Institute of Microsystem and Information Technology, Chinese Academy of Sciences, Shanghai 200050, China}
\affiliation{Center of Materials Science and Optoelectronics Engineering, University of Chinese Academy of Sciences, Beijing 100049, China}
\affiliation{School of Physical Science and Technology, ShanghaiTech University, Shanghai 201210, China}

\author{X. B. Shi}
\thanks{Equal contributions}
\affiliation{State Key Laboratory of Advanced Welding and Joining, Harbin Institute of Technology, Shenzhen 518055, China}
\affiliation{Flexible Printed Electronics Technology Center, Harbin Institute of Technology, Shenzhen 518055, China}

\author{G. N. Zhang}
\affiliation{School of Physical Science and Technology, ShanghaiTech University, Shanghai 201210, China}

\author{Z. T. Liu}
\affiliation{Center for Excellence in Superconducting Electronics, State Key Laboratory of Functional Materials for Informatics, Shanghai Institute of Microsystem and Information Technology, Chinese Academy of Sciences, Shanghai 200050, China}

\author{Soohyun Cho}
\affiliation{Center for Excellence in Superconducting Electronics, State Key Laboratory of Functional Materials for Informatics, Shanghai Institute of Microsystem and Information Technology, Chinese Academy of Sciences, Shanghai 200050, China}

\author{Z. C. Jiang}
\affiliation{Center for Excellence in Superconducting Electronics, State Key Laboratory of Functional Materials for Informatics, Shanghai Institute of Microsystem and Information Technology, Chinese Academy of Sciences, Shanghai 200050, China}

\author{Z. H. Liu}
\affiliation{Center for Excellence in Superconducting Electronics, State Key Laboratory of Functional Materials for Informatics, Shanghai Institute of Microsystem and Information Technology, Chinese Academy of Sciences, Shanghai 200050, China}
\affiliation{Center of Materials Science and Optoelectronics Engineering, University of Chinese Academy of Sciences, Beijing 100049, China}

\author{J. S. Liu}
\email{jishanliu@mail.sim.ac.cn}
\affiliation{Center for Excellence in Superconducting Electronics, State Key Laboratory of Functional Materials for Informatics, Shanghai Institute of Microsystem and Information Technology, Chinese Academy of Sciences, Shanghai 200050, China}

\author{X. L. Lu}
\affiliation{Center for Excellence in Superconducting Electronics, State Key Laboratory of Functional Materials for Informatics, Shanghai Institute of Microsystem and Information Technology, Chinese Academy of Sciences, Shanghai 200050, China}

\author{Y. C. Yang}
\affiliation{Center for Excellence in Superconducting Electronics, State Key Laboratory of Functional Materials for Informatics, Shanghai Institute of Microsystem and Information Technology, Chinese Academy of Sciences, Shanghai 200050, China}

\author{W. Xia}
\affiliation
{School of Physical Science and Technology, ShanghaiTech University, Shanghai 201210, China}
\affiliation{\mbox{ShanghaiTech Laboratory for Topological Physics, ShanghaiTech University, Shanghai 201210, China}}

\author{W. W. Zhao}
\affiliation{State Key Laboratory of Advanced Welding and Joining, Harbin Institute of Technology, Shenzhen 518055, China}
\affiliation{Flexible Printed Electronics Technology Center, Harbin Institute of Technology, Shenzhen 518055, China}

\author{Y. F. Guo}
\email{guoyf@shanghaitech.edu.cn}
\affiliation{School of Physical Science and Technology, ShanghaiTech University, Shanghai 201210, China}

\author{D. W. Shen}
\email{dwshen@mail.sim.ac.cn}
\affiliation{Center for Excellence in Superconducting Electronics, State Key Laboratory of Functional Materials for Informatics, Shanghai Institute of Microsystem and Information Technology, Chinese Academy of Sciences, Shanghai 200050, China}
\affiliation{Center of Materials Science and Optoelectronics Engineering, University of Chinese Academy of Sciences, Beijing 100049, China}

\begin{abstract}

The signatures of topological superconductivity (TSC) in the superconducting materials with topological nontrivial states prompt intensive researches recently. Utilizing high-resolution angle-resolved photoemission spectroscopy and first-principles calculations, we demonstrate multiple Dirac fermions and surface states in superconductor BaSn$_3$ with a critical transition temperature of about 4.4 K. We predict and then unveil the existence of two pairs of type-\uppercase\expandafter{\romannumeral1} topological Dirac fermions residing on the rotational axis. Type-\uppercase\expandafter{\romannumeral2} Dirac fermions protected by screw axis are confirmed in the same compound. Further calculation for the spin helical texture of the observed surface states originating from the Dirac fermions give an opportunity for realization of TSC in one single material. Hosting multiple Dirac fermions and topological surface states, the intrinsic superconductor BaSn$_3$ is expected to be a new platform for further investigation of the topological quantum materials as well as TSC.

\end{abstract}

\maketitle

\section{Introduction}

The long-sought topological superconductivity (TSC) has attracted vast attention and got boomingly developed in the last decade, which were spurred by the potential realization of Majorana fermions~\cite{qi2011topological,alicea2012new,leijnse2012introduction,beenakker2013search,stanescu2013majorana,aguado2017majorana,sato2017topological}. In this context, there has been an explosion of proposals to search for novel materials hosting TSC, including intrinsic odd-parity superconductors~\cite{mackenzie2003superconductivity,fu2010odd,sato2010topological,hsieh2012majorana} and heterostructures taking advantage of the superconducting proximity effect of spin helical surface states in topological insulators (TIs)~\cite{fu2008superconducting} or spin polarization in Rashba semiconductors~\cite{alicea2010majorana,sau2010generic,lutchyn2010majorana,lutchyn2018majorana,frolov2020topological} to conventional \emph{s}-wave pairing in superconductors. However, for the former strategy, the paring symmetry of devised candidates for \emph{p}-wave superconductors still remains elusive~\cite{kashiwaya2011edge,jang2011observation,hor2010superconductivity,wray2010observation,sasaki2011topological,matano2016spin,yonezawa2017thermodynamic,liu2015superconductivity}. As for the case of heterostructures, the utilization of spin-triplet pairing feasible at interfaces was hindered by the intricacy of fabrication and the disturbance of interface physics~\cite{williams2012unconventional,wang2012coexistence,wang2013fully,cho2013symmetry,oostinga2013josephson,finck2014phase,hart2014induced,mourik2012signatures,rokhinson2012fractional,das2012zero,deng2012anomalous,churchill2013superconductor,finck2013anomalous,lee2014spin,nadj2014observation,albrecht2016exponential,deng2016majorana,zhang2017ballistic,zhang2018quantized,article}.
Recent researches reported evidences of \emph{s}-wave superconductors hosting TI states with spin polarized surface states and Majorana zero mode in the vortex core, suggestive of a new way to realize TSC in one single material~\cite{zhang2018observation,wang2018evidence,gray2019evidence,machida2019zero,kong2019half,wang2020evidence,zhu2020nearly}.
Besides, \emph{s}-wave superconductors with topological Dirac semimetal states have as well been proposed as candidates for TSC~\cite{PhysRevB.100.094520,zhang2019multiple}, taking into account of the helical surface states around the Fermi level ($E_F$) brought by Dirac bands. Experimentally, the possible unconventional superconductivity has been reported to be induced by point contact~\cite{aggarwal2016unconventional,wang2016observation} or high pressure~\cite{he2016pressure} in the prototype Dirac semimetal Cd$_3$As$_2$. However, such extreme external conditions would prohibit further investigation and application of TSC, and searching for intrinsic superconducting topological Dirac semimetal is highly desired.

Recently, superconducting compounds ASn$_3$ (A=Ca or Ba) were predicted to host topological semimetal states~\cite{luo2015superconductivity,zhu2019emergence,siddiquee2021fermi,zhang2020haas,guechi2021structural}. Studies on the topology in the weak coupling superconductor CaSn$_3$ propose that it hosts topological nodal-line semimetal states in the absent of spin-orbital coupling (SOC) and would evolve to Weyl nodes with SOC~\cite{luo2015superconductivity,zhu2019emergence,siddiquee2021fermi}. While, its relatively complicated band structure impedes the direct investigation of the nontrivial band topology therein. Fascinatingly, first-principles calculations suggest that the electronic band structure of BaSn$_3$ harbors multiple simple topological Dirac semimetal states, and Type-\uppercase\expandafter{\romannumeral2} bulk superconductivity has been experimentally confirmed with an onset superconducting temperature \emph{T$_c^{onset}$} $\sim$ 4.4 K~\cite{fassler1997basn3,zhang2020haas}. Moreover, de Haas-van Alphen (dHvA) measurements on this compounds have provided evidence on the possible nontrivial Berry phase associated with the proposed type-\uppercase\expandafter{\romannumeral2} Dirac point around $E_F$, making it a good candidate to realize TSC and Majorana fermions~\cite{zhang2020haas}. However, to date, the direct confirmation of proposed multiply Dirac fermions in the low-lying electronic structure is still lack. 

\begin{figure*}[ht]
\centering
\includegraphics[width=\columnwidth]{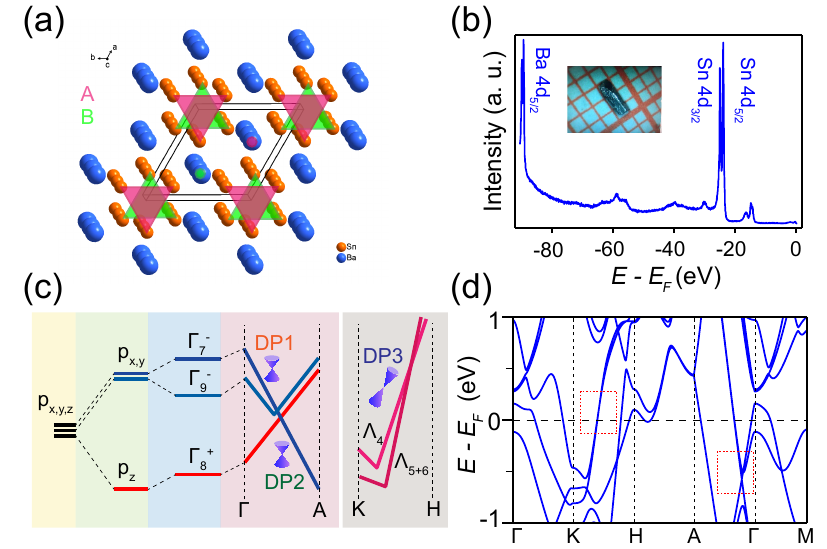}
\caption{(a) Crystal structure of BaSn$_3$. (b) core-level photoemission spectrum clearly shows the characteristic Ba 4\emph{d} and Sn 4\emph{d} peaks. Inset: Photograph of the high-quality BaSn$_3$ single crystal with shiny surface ued for ARPES measurement. (c) Schematic illustration of \emph{p}-orbital energy level evolution to form the type-\uppercase\expandafter{\romannumeral1} DPs. Right column: The band structure of type-\uppercase\expandafter{\romannumeral2} DPs. The symmetry of energy levels is labeled by the IRs. (d) The calculated band structure of BaSn$_3$.}
\label{crystal structure}
\end{figure*}

In this work, we predicted and then directly identified the multiple Dirac fermions in the superconducting BaSn$_3$ by employing first-principles calculations and high-resolution angle-resolved photoemission spectroscopy (ARPES). Our ARPES result on the unique (100) cleavage plane verified the existence of two pairs of type-\uppercase\expandafter{\romannumeral1} three-dimensional Dirac points (DPs) residing on the rotational axis. In the same compound, we discovered type-\uppercase\expandafter{\romannumeral2} DPs protected by the screw axis. Moreover, we deduced that the observed surface states with predicted spin helical texture could induce the possible TSC in one single material. Our experimental result is complementary to the previous transport measurements, and consolidates the evidence on the nontrivial topological bands and two types of Dirac fermions in BaSn$_3$. Our findings uncover the topological electronic states in superconducting BaSn$_3$, which might provide a new platform for further studies on the topological quantum materials as well as TSC.

\section{Materials and methods}

High-quality single crystals of BaSn$_3$ were grown with a self flux method. Details on the sample growth and characterization can be found elsewhere~\cite{zhang2020haas}. High-resolution ARPES measurements were performed at 03U beam line of Shanghai Synchrotron Radiation Facility (SSRF) and 13U beam line of National Synchrotron Radiation Laboratory (NSRL), respectively. Both end stations are equipped with Scienta Omicron DA30 electron analyzers. Considering the sensitivity of BaSn$_3$ to water in the air, all samples were meticulously selected and prepared in the glove box first, and then cleaved along both the (001) and (100) cleavage planes. All ARPES data were taken at 15 K in an ultrahigh vacuum better than 8.0 $\times$ 10$^{-11}$ mbar. The angular and the energy resolutions were set to 0.2$^\circ$ and 6$\sim$20 meV (depending on the incident photon energies), respectively.

First-principles calculations were performed within the framework of the projector augmented wave (PAW) method and selected the generalized gradient approximation (GGA) with Perdew-Burke-Ernzerhof (PBE) type, as encoded in the Vienna Ab initio Simulation Package (VASP). A kinetic energy cutoff of 500 eV and a $\Gamma$-centred k mesh of 6 $\times$ 6 $\times$ 9 were utilized for all calculations. During self-consistent convergence and structural relaxation, the energy and force difference criterion were defined as 10$^{-6}$ eV and 0.01 eV/$\AA$. SOC was considered in a self-consistent manner. The WANNIER90 package was adopted to construct Wannier functions from the first-principles results. Topological properties calculations were carried out by using the WANNIERTOOLS code.

\begin{figure*}[htbp]
\includegraphics[width=17cm]{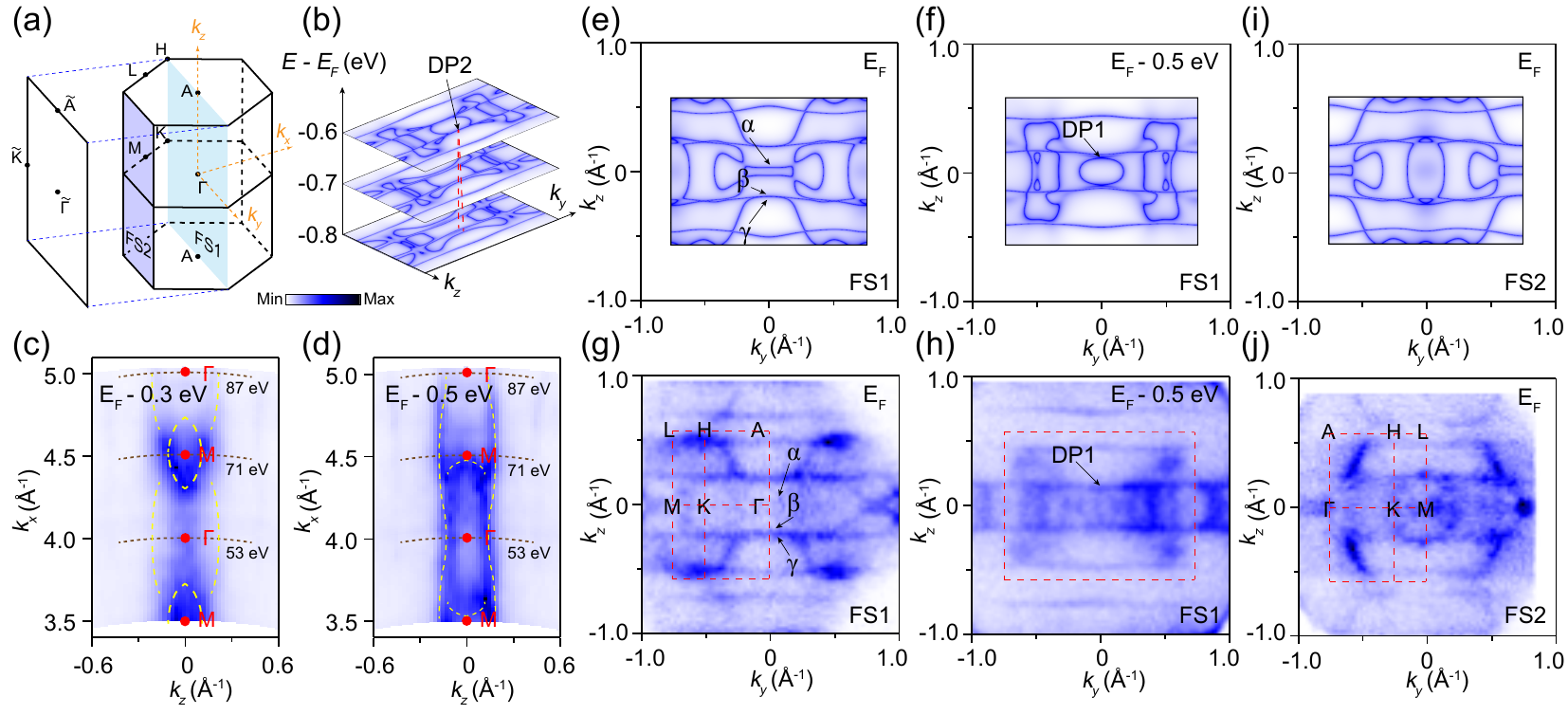}
\caption{(a) Bulk BZ with high-symmetry points. Two planes labeled as FS1 and FS2 indicate the \emph{k$_x$} = 0 and \emph{k$_x$} = $\pi$, respectively. (b) Stack of CECs in the \emph{k$_y$-k$_z$} plane showing the lower part of Dirac cone. (c),(d) Photoemission intensity plots in the \emph{k$_x$-k$_z$} plane at $E_F$ - 0.3 eV and $E_F$ - 0.5 eV, respectively. (e),(f) Calculated bulk FSs in the high-symmetry plane FS1 at $E_F$ = 0 and -0.5 eV, respectively. (g),(h) Photoemission intensity plots at $E_F$ = 0 and -0.5 eV, showing CECs on FS1 indicated in (a). (i) Calculated bulk FSs in the high-symmetry plane FS2 at $E_F$. (j) Photoemission intensity plots at $E_F$ in FS2. The data were recorded on the (100) cleavage plane with linearly horizontal polarized photons.  (g)-(h) were taken at 87 eV. (j) was taken at 71 eV.}
\label{3D electronic structure}
\end{figure*}

\section{Results and Discussion}

BaSn$_3$ crystallizes in the hexagonal Ni$_3$Sn structure with the space group \emph{$P6_{3}mmc$}~(No. 194)~\cite{fassler1997basn3}. Two hexagonal layers containing Ba and Sn atoms are stacked in an A-B-A-B sequence to form a close-packed structure as schematically illustrated in Fig.~\ref{crystal structure}(a). Thus, the (100) cleavage plane is easier to be obtained than the (001) one [Fig. 1(a)], which provides a good opportunity to directly observe both types of DPs. Previous single-crystal X-ray diffraction (XRD) and energy dispersive spectroscopy (EDS) measurements have confirmed the high crystalline quality and accurate stoichiometric proportion of our samples~\cite{zhang2020haas}. Furthermore, our core-level photoemission measurements further confirm the chemical composition of BaSn$_3$ [Fig.~\ref{crystal structure}(b)], showing sharp Ba 4\emph{d} and Sn 4\emph{d} orbital peaks. We note that our samples show a superconducting transition temperature of 4.4 K and behave as a type \uppercase\expandafter{\romannumeral2} superconductor~\cite{zhang2020haas}. After cleaved in the air, the sample shows typical flat and shining surface as illustrated in the inset of Fig.~\ref{crystal structure}(b).

The crystal structure of BaSn$_3$ respects both the inversion and \emph{C$_3$} rotation symmetries, protecting the predicted topological DPs~\cite{fassler1997basn3,zhang2020haas}. Here, the nontrivial topological band structure of BaSn$_3$ originates from the band inversion within the \emph{p} orbitals of Sn as shown in Fig.~\ref{crystal structure}(c). When the crystal-field splitting is under consideration, the \emph{p} orbitals split into \emph{p$_{x,y}$} and \emph{p$_z$} possessing different energies. When the SOC participates in it, the in-plane bands from \emph{p$_{x,y}$} orbital further splits into bands which belongs to $\varGamma_7^-$ and $\varGamma_9^-$ irreducible representations (IR) and cross with each other. Meanwhile, the $\varGamma_7^-$ band crosses with the $\varGamma_8^+$ band originating from the \emph{p$_z$} orbital. Due to different IRs of these crossing bands, two type-\uppercase\expandafter{\romannumeral1} DPs marked as DP1 and DP2 form along $\varGamma-\emph{A}$ direction. Protected by the inversion and time reversal symmetry, all bands in BaSn$_3$ is doubly degenerate. Therefore, the DP1 and DP2 are fourfold degenerate and protected by the additional \emph{C$_3$} rotation symmetry to prevent the gap from opening due to the SOC effect. More interestingly, along the $\emph{K}-\emph{H}$ direction, there is an unavoidable crossing between two bands forming the strongly tilted type-\uppercase\expandafter{\romannumeral2} Dirac fermion labeled as DP3 as illustrated in the right column of Fig.~\ref{crystal structure}(c). The symmetry analysis indicates that two crossing bands belong to $\varLambda_4$ and $\varLambda_{5+6}$, respectively. Fig.~\ref{crystal structure}(d) shows total band dispersions along high-symmetry directions with SOC. The electronic structure near $E_F$ is mainly dominated by Sn 5\emph{p} orbitals [Details in Note 1 in Supplemental Material (SM)~\cite{si}]. We highlight the bands related to topological DPs by red boxes.

Experimentally, we first investigate the overall electronic structure of BaSn$_3$. The three-dimensional Brillouin zone (BZ) of BaSn$_3$ is illustrated in Fig.~\ref{3D electronic structure}(a), in which plotted black dots indicate high-symmetry points. The directions of \emph{k$_x$}, \emph{k$_y$} and \emph{k$_z$} are defined to parallel to $\varGamma-\emph{M}$, $\varGamma-\emph{K}$ and $\varGamma-\emph{A}$, respectively. To trace the evolution of the bulk electronic structures and determine the high symmetry surface, systematic photon energy dependent measurements were performed (photon energy ranged from 41 eV to 91 eV). Here, we discuss the \emph{k$_x$}-momentum determination. Since the surface of the measured sample breaks the translational symmetry along out-of-plane direction, the perpendicular momentum components of the photonelectron (\emph{k$_\perp^f$}) and the initial electron (\emph{k$_\perp^i$}) are not equivalent. Although we can not get the out-of-plane momentum directly as the in-plane momentum (\emph{k$_z$} and \emph{k$_y$}), it is helpful to determine the \emph{k$_x$} momentum based on the nearly free-electron approximation for final states~\cite{lv2019angle},
 \begin{equation}
   k_{\operatorname{\perp}}^{\operatorname{i}} = \sqrt{2m(E_{\operatorname{kin}}\cos^{\operatorname{2}}\theta)+V_{\operatorname{0}}}/\hbar
\end{equation}
where V$_0$ is the inner potential which is a constant and can be defined by fitting the periodicity of different photon energy measurement. As shown in Figs.~\ref{3D electronic structure}(c) and \ref{3D electronic structure}(d), the clear periodic patterns were observed in both \emph{k$_x$-k$_z$} planes at $E_F$ - 0.3 eV and $E_F$ - 0.5 eV. We thus got the inner potential V$_0$ = 13 eV in BaSn$_3$ from the \emph{k$_x$} dependent photoemission intensity plots. Thus, we can determine the exact value of \emph{k$_\perp$} and corresponding high symmetry points and surfaces.

The photoemission intensity maps of constant energy contours (CECs) at $E_F$ on in-plane FS1 and FS2 [Figs.~\ref{3D electronic structure}(g) and \ref{3D electronic structure}(j)] indicate the Fermi surfaces (FSs) on the \emph{k$_x$} = 0 and \emph{k$_x$} = $\pi$, respectively, as schematically illustrated in Fig.~\ref{3D electronic structure}(a), which are in perfect agreement with calculations [Figs.~\ref{3D electronic structure}(e) and \ref{3D electronic structure}(i)]. From the detailed investigation of CECs on FS1 at different binding energy, we could recognize both type-\uppercase\expandafter{\romannumeral1} Dirac points DP1 and DP2. One holelike band $\alpha$ belonging to $\varGamma_7^-$ and two electronlike bands $\beta$, $\gamma$ belonging to $\varGamma_9^-$, $\varGamma_8^+$ could be identified both in the calculation [Fig.~\ref{3D electronic structure}(e)] and experimental result [Fig.~\ref{3D electronic structure}(g)] near $\varGamma$ at $E_F$. With the increase of binding energy, the hole pocket from $\alpha$ band near $\varGamma$ extends, while the electron one from $\beta$ band shrinks. More details on CECs at different binding energies can be found in Note 2 of SM~\cite{si}. As shown in the Fig.~\ref{3D electronic structure}(h), the above two pockets finally overlap at $E_F$ - 0.5 eV where form the DP1, in a remarkable agreement with the theory [Fig.~\ref{3D electronic structure}(f)]. Meanwhile, from the stack of calculated CECs results at 0.6, 0.7 and 0.8 eV below $E_F$ [Fig.~\ref{3D electronic structure}(b)], we could identify the evolution of the lower part of the Dirac cone from DP2 located at $\sim$0.6 eV below $E_F$.

\begin{figure*}[htbp]
\includegraphics[width=17cm]{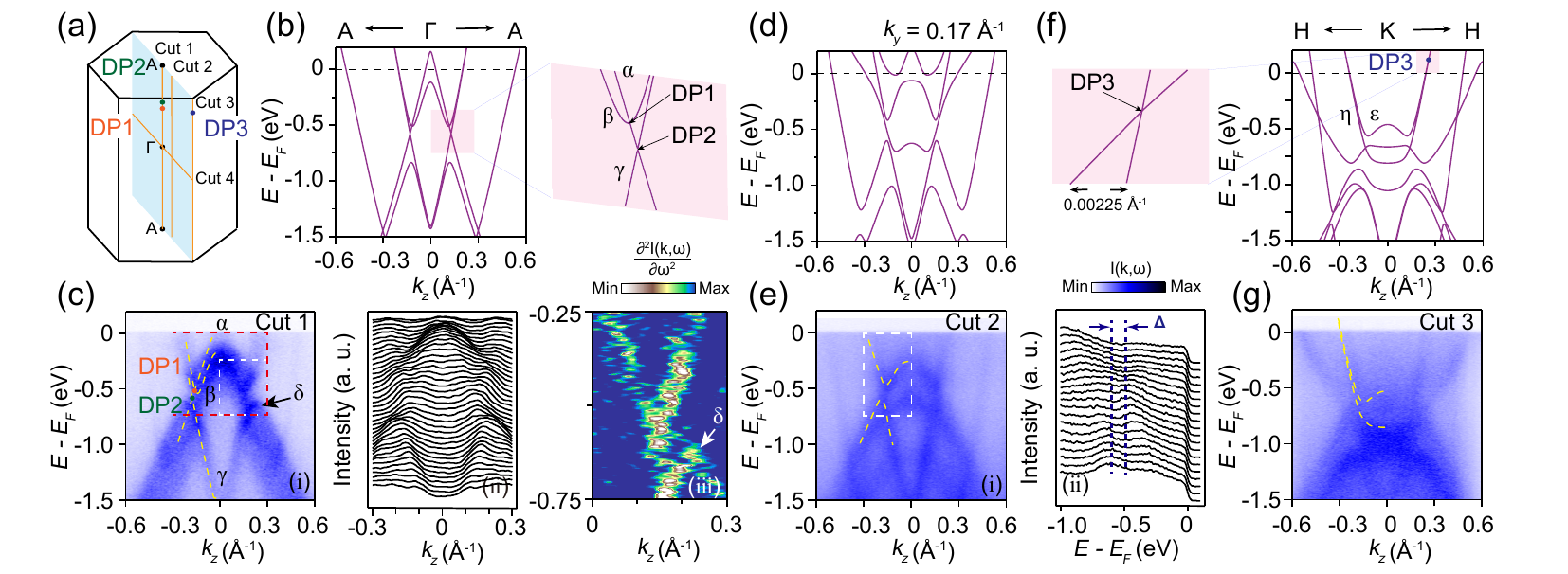}
\caption{(a) Bulk BZ with orange lines indicating the location of cuts 1 to 4. (b) Calculated band structure along $\varGamma-\emph{A}$ high-symmetry direction with SOC. Right column: the enlarged image of the red box, displaying two pair of type-\uppercase\expandafter{\romannumeral1} DPs. (b) ($\romannumeral1$) Photoemission intensity plots of band dispersions along cut 1. ($\romannumeral2$) MDCs of region in the white box. ($\romannumeral3$) The enlarged second derivative intensity plot of region in the white box. (d) Calculated band structure along cut 2 with SOC. (e) ($\romannumeral1$) Photoemission intensity plots of band dispersions along cut 2. ($\romannumeral2$) MDCs of region in the white box. (f) Calculated band structure along $\emph{K}-\emph{H}$ high-symmetry direction with SOC. Left column: the enlarged broadening image of the red box, displaying tilted type-\uppercase\expandafter{\romannumeral2} DP. (g) Photoemission intensity plots of band dispersions along cut 3. The ARPES data were recorded on the (100) cleavage surface with 53 eV in (c) and 28 eV in (e),(g) under linearly horizontal polarization.}
\label{Dirac cone}
\end{figure*}

Next, we focus on the FS1 plane to identify the two types of Dirac fermions along $\varGamma-\emph{A}$ and $\emph{K}-\emph{H}$ high symmetry directions [Cut 1 and Cut 3 in Fig.~\ref{Dirac cone}(a)]. The band calculation along $\varGamma-\emph{A}$ direction indicates that DP1 is located at $\sim$0.5 eV below $E_F$ where $\alpha$ and $\beta$ bands cross, while the DP2 is located at $\sim$0.6 eV below $E_F$ where $\alpha$ and $\gamma$ bands cross [Fig.~\ref{Dirac cone}(b)]. To search for the theoretically proposed DP1 and DP2, we measured the band dispersions along the $\varGamma-\emph{A}$, namely, Cut 1 as shown in Fig.~\ref{Dirac cone}(c). The ARPES intensity plot exhibits three bands and two crossing points, consistent with the theoretical prediction. Since BaSn$_3$ respects both the inversion and time-reversal symmetries, these two intersections between $\alpha$/$\beta$, and  $\alpha$/$\gamma$ are both fourfold degenerate. The additional $C_3$ symmetry protects Dirac fermions from gapping out, finally forming two bulk topological nontrivial type-\uppercase\expandafter{\romannumeral1} DPs marked as DP1 and DP2. To further locate these two DPs, we display the enlarged second derivative intensity plot [Fig.~\ref{Dirac cone}(c)($\romannumeral2$)] and momentum distribution curves (MDCs) [Fig.~\ref{Dirac cone}(c)($\romannumeral3$)] of the region in the white box. Here, the calculated band structure has been appended on top of the experimental data in the enlarged second derivative plot, from which we can identified the DP1 situated at $E_F$ - 0.5 eV, and DP2 at $E_F$ - 0.6 eV. Thus, we directly demonstrate and locate both predicted type-\uppercase\expandafter{\romannumeral1} DPs in BaSn$_3$.

To further trace the band structure related to DP1 and DP2, we measured detailed band dispersions on the FS1 plane parallel to $\varGamma-\emph{A}$ as illustrated in Fig.~\ref{Dirac cone}(a). The experimental band dispersions at \emph{k$_y$} = 0 $\AA^{-1}$ exhibit one hole-like band $\alpha$, one ``W''-shaped band $\beta$ and one ``V''-shaped band $\gamma$ [Fig.~\ref{Dirac cone}(c) ($\romannumeral2$)], with two crossing points DP1 and DP2, respectively. Note that the ``M''-shaped band $\delta$ which is caused by the electronic intensity from other high-symmetry direction parallel to $\varGamma-\emph{A}$ direction does not participate in the formation of DPs according to our slab calculation [see details in Note 3 of SM~\cite{si}]. Since all DPs along $\varGamma-\emph{A}$ are protected by the $C_3$ rotation symmetry, upon sliding the cut to \emph{k$_y$} = 0.17 $\AA^{-1}$ [Cut 2 shown in Fig.~\ref{Dirac cone}(a)], which deviates from high symmetry lines, the gaps near two DPs are turned on [Fig.~\ref{Dirac cone}(e)], which is consistent with the calculated result [Fig.~\ref{Dirac cone}(d)]. Furthermore, the corresponding MDCs for the enlarged region can further reveal the band gap, as shown in Fig.~\ref{Dirac cone}(e)($\romannumeral2$).

In addition to the two type-\uppercase\expandafter{\romannumeral1} DPs along $\varGamma-\emph{A}$, we discovered the evidence for the existence of type-\uppercase\expandafter{\romannumeral2} DP along $\emph{K}-\emph{H}$. Upon further sliding the cut to \emph{k$_y$} = 0.5 $\AA^{-1}$ [Cut 3 in Fig.~\ref{Dirac cone}(a)], which precisely slices the $\emph{K}-\emph{H}$ direction, the band structures exhibit one ``W''-shaped band $\epsilon$ and one ``U''-shaped band $\eta$ [Fig.~\ref{Dirac cone}(g)], in remarkable line with the calculation [Fig.~\ref{Dirac cone}(f)]. Since the $\epsilon$ and $\eta$ bands belong to two different IRs $\varLambda_4$ and $\varLambda_{5+6}$, respectively, the unavoidable crossing between them will occur above the $E_F$ as calculated in Fig.~\ref{Dirac cone}(f). We note that from the the enlarged image here, the $\epsilon$ and $\eta$ bands are only 0.00225 $\AA^{-1}$ apart at $E_F$, beyond the momentum resolution of our photoemission experiments. Thus, we can not separate the $\epsilon$ and $\eta$ bands at $E_F$. The type-\uppercase\expandafter{\romannumeral2} DPs strongly tilting along the $\emph{K}-\emph{H}$ are protected by the screw axes, inversion and time-reversal symmetry.

\begin{figure*}[htbp]
\centering
\includegraphics[width=17cm]{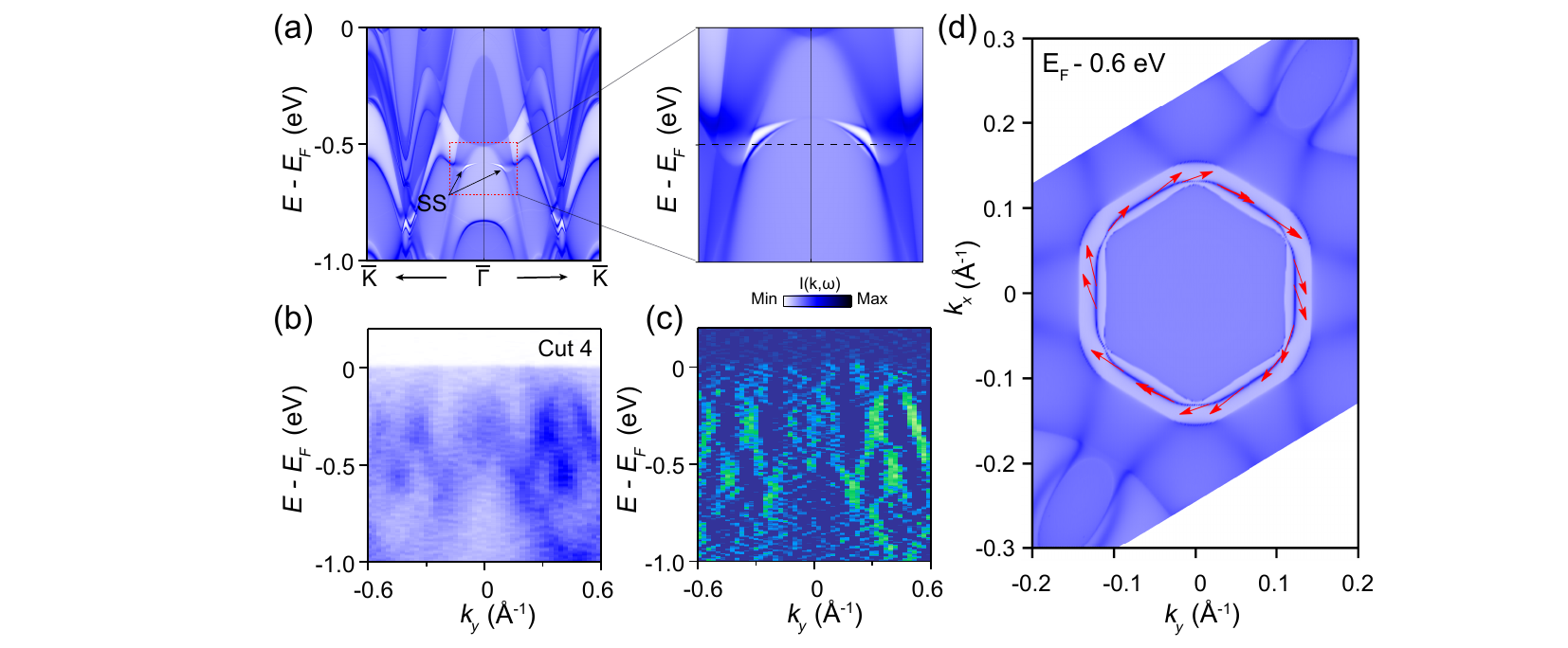}
\caption{(a) Calculated topological surface states along $\bar{\varGamma}-\bar{\emph{K}}$ high-symmetry direction. Right column: the enlarged calculated result in the red box. (b) Photoemission intensity plots of band dispersions along $\varGamma-\emph{K}$ high-symmetry direction. The surface states labeled as SS are marked by black arrow. (c) The MDCs of the white box in (b). The ARPES data were taken with 87 eV under linearly horizontal polarization. (d) Calculated spin texture at 0.6 eV below $E_F$, displaying the spin-momentum locking texture.}
\label{surface states}
\end{figure*}

In Fig.~\ref{surface states}, we investigated the topological surface states originating from the observed DP2 and the spin polarized texture. The calculated surface states from the DP2 shown in Fig.~\ref{surface states}(a) were marked as SS by black arrows along the $\bar{\varGamma}-\bar{\emph{K}}$ direction. The photoemission intensity plot along $\varGamma-\emph{K}$ high symmetry direction [Cut 4 in Fig.~\ref{Dirac cone}(a)] shown the surface states at $\sim$0.6 eV below $E_F$, which is consistent with the calculated result. From the second derivative intensity plot in Fig.~\ref{surface states}(c), we could recognize the surface states more clearly. To further study the spin texture of these surface states, we calculated the constant-energy plot of spin texture at $E_F$ - 0.6 eV. These red arrows in Fig.~\ref{surface states}(d) show the spin direction of electrons, indicating the characteristic spin-momentum locking texture of these surface states.

The coexistence of bulk DPs and spin-helical surface states in superconductor BaSn$_3$ resembles that of iron-based superconductor [e.g., Fe(Te,Se), Li(Fe,Co)As]~\cite{PhysRevB.100.094520,zhang2019multiple}. Taking advantage of the proximity effect in momentum space, the possible helical surface states could be induced to get the superconducting order parameter, in which the superconducting gap could be opened and realize the TSC in one single material. Furthermore, the hole doping will lower the $E_F$ and make the DPs and surface states closer to the $E_F$. Then the FS with mixing parities could form. Therefore, the superconductor BaSn$_3$ with $T_c$ $\sim$ 4.4 K could be an exciting platform to enrich the potential TSC family.

\section{Conclusion}
In summary, we have identified the multiple Dirac fermions and related surface states with predicted spin-momentum locking texture in the superconductor BaSn$_3$ by ARPES measurements in association with first-principles calculations. The high resolution ARPES results recognize the dual type-\uppercase\expandafter{\romannumeral1} DPs and provide evidence for the type-\uppercase\expandafter{\romannumeral2} DP complementary to the previous transport measurement. The topological Dirac fermions and  surface states make BaSn$_3$ a good platform for exploring the topological quantum materials. Our findings are also beneficial for researches on TSCs.

\begin{acknowledgments}

This work was supported by the National Key R$\&$D Program of the MOST of China (Grant No. 2016YFA0300204), the National Science Foundation of China (Grant Nos. U2032208, 11874264), and the Natural Science Foundation of Shanghai (Grant No. 14ZR1447600). Y. F. Guo acknowledges the starting grant of ShanghaiTech University and the Program for Professor of Special Appointment (Shanghai Eastern Scholar). Part of this research used Beamline 03U of the Shanghai Synchrotron Radiation Facility, which is supported by ME$^2$ project under contract No. 11227902 from National Natural Science Foundation of China. ZJW was supported by the National Nature Science Foundation of China (Grant No. 11974395), the Strategic Priority Research Program of Chinese Academy of Sciences (Grant No. XDB33000000), and the Center for Materials Genome. The authors also thank the support from Analytical Instrumentation Center (\#SPST-AIC10112914), SPST, ShanghaiTech University.

\end{acknowledgments}

\bibliographystyle{apsrev4-1}

%


\end{document}